\documentstyle[12pt]{article}
\input epsf.tex
\begin{document}\bigskip
\hskip 3.7in\vbox{\baselineskip12pt
\hbox{NSF-ITP-97-061}
\hbox{Imperial/TP/96-97/49}
\hbox{hep-th/9706072}}

\bigskip\bigskip

\centerline{\large \bf Higher Order Graviton Scattering}
\centerline{\large \bf in M(atrix) Theory}

\bigskip\bigskip

\centerline{{\bf Katrin Becker$^1$, Melanie Becker$^2$,}
}
\centerline{{\bf
Joseph Polchinski$^1$} and {\bf Arkady Tseytlin$^3$}
}

\bigskip

$^1${\it Institute for Theoretical Physics,
University of California, Santa Barbara, CA 93106-4030}

\bigskip

$^2${\it Department of Physics,
University of California, Santa Barbara, CA 93106}

\bigskip

$^3${\it  Blackett Laboratory,
Imperial College,  London,  SW7 2BZ and Lebedev Institute, Moscow}

\bigskip

\begin{abstract}
\baselineskip=16pt
In matrix theory the effective action for graviton-graviton scattering is a
double expansion in the relative velocity and inverse separation.  We
discuss the systematics of this expansion and subject matrix theory to a new
test.  Low energy supergravity predicts the coefficient of the $v^6/r^{14}$
term, a two-loop effect, in agreement with explicit matrix model calculation.
\end{abstract}
\newpage
\baselineskip=18pt

Matrix theory is a remarkable proposal for the fundamental degrees of
freedom and their Hamiltonian.  In the original paper~\cite{bfss}, one of
the principal tests was a successful comparison of graviton-graviton
scattering in the matrix theory and in eleven-dimensional supergravity.
In subsequent work this has been extended to scattering of extended
objects~\cite{morescat,bercor} and scattering with nonzero momentum
transfer $q_{11}$~\cite{keleven}.  This work has dealt with the leading
term both at low velocity and long distance, for example $v^4/r^7$ in
graviton-graviton scattering at $q_{11}=0$.  Matrix theory predicts a
series of corrections both in $v$ and $r$, and if it is correct these must
all be understood in eleven-dimensional terms.

Two of the present authors~\cite{beckers} have recently reported on the
$v^4/r^{10}$ term at $q_{11}= 0$, a two-loop matrix theory effect.  It
vanishes as required by the matrix theory conjecture.  In this note we
would like to develop some of the systematics of the double expansion in
$v$ and $r$ for graviton-graviton scattering at $q_{11}=0$ and to report
on a new test of matrix theory.  We observe that eleven-dimensional
supergravity predicts the coefficient of the $v^6/r^{14}$ term.  In matrix
theory this is a two-loop effect, and by an extension of the
calculation~\cite{beckers} we find agreement.\footnote
{A recent paper by Ganor, Gopakumar, and Ramgoolam~\cite{ggr} considers
scattering of a graviton from an $R^8/Z_2$ fixed point, finding a
discrepancy at two loops.  The extension of matrix theory to such
less symmetric backgrounds is an open issue.  These authors also 
discuss the general form of higher order matrix theory amplitudes.}

Higher velocity corrections have recently been considered in
ref.~\cite{ballar}.  In that work there appeared to be a mismatch between
the supergravity and matrix theory amplitudes.  However, as noted by those
authors, the mismatch is subleading in the large-$N$ expansion.  To make
the comparison one must therefore have a precise understanding of the
meaning of the finite-$N$ matrix theory.  Happily, this has recently been
supplied in an important paper by Susskind~\cite{dlqm} (see
also~\cite{dvv}).  Finite
$N$ is to be identified with compactification of a null direction
(henceforth the $-$ direction), not a spacelike direction.  We
will see that the velocity expansion at fixed
$p_-$ is simpler than in ref.~\cite{ballar}.

Let us consider first the matrix theory perturbation expansion.  The
bosonic part of the matrix theory action is~\cite{bfss}
\begin{equation}
S = \int d\tau\,{\rm Tr}\left( \frac{1}{2R} D_\tau X^i D_\tau X^i +
\frac{M^6 R}{4}
[X^i,X^j]^2 \right)\ ,
\end{equation}
where $R$ is the radius of eleventh dimension and
$M$ the eleven-dimensional Planck mass up to a convention-dependent
numerical coefficient; the signs are appropriate for Hermitean $X$.
By rescaling $\tau = u/R$ and $X^i =
y^i/M^3$, the action becomes
\begin{equation}
S = \frac{1}{M^6} \int du\,{\rm Tr}\left( \frac{1}{2} D_u y^i D_u y^i +
\frac{1}{4} [y^i,y^j]^2 \right)\ .
\end{equation}
It follows that $M^6$ is the loop-counting parameter, and that the
effective action at $L$ loops is of the form
\begin{equation}
S_L\ =\ M^{6L-6} \int du\, f_L(y^i, D_u)\
=\ RM^{6L-6} \int d\tau\, f_L(M^3 X^i, R^{-1} D_\tau)\ .
\end{equation}
Finally we have dimensional analysis: $f_L$ must have units of (length)$^{6L
- 8}$.  For the leading low energy effective action, depending on the
velocity~$v^i = D_\tau X^i$ but not the acceleration, this becomes
\begin{equation}
S_L = RM^{6-3L} \int d\tau\, r^{4-3L}
g_L\left( \frac{X^i}{r}, \frac{v^i}{R M^3 r^2}\right)
\end{equation}
where $r^2 = X^i X^i$.
Let us write out the first few terms in the expansion of the effective
Lagrangian, indicating the dependence on $v$ and $r$ but suppressing the
dependence on
$M$, $R$, and $X^i / r$:\footnote
{The systematics of this expansion have also been considered by W. Fischler
and L. Susskind~\cite{susspriv}.}
\begin{equation}
\begin{array}{cccccccccc}
\displaystyle
{\cal L}_0 &=& c_{00} v^2 &&&&&&& \\
{\cal L}_1 &=&  && \displaystyle c_{11} \frac{v^4}{r^7} & +
&\displaystyle c_{12} \frac{v^6}{r^{11}}& + &
\displaystyle c_{13} \frac{v^8}{r^{15}} & +\ \ldots \\[8pt]
{\cal L}_2 &=&  && \displaystyle c_{21} \frac{v^4}{r^{10}} & +
&\displaystyle c_{22} \frac{v^6}{r^{14}}& + &
c_{23} \displaystyle\frac{v^8}{r^{18}} & +\ \ldots \\[8pt]
{\cal L}_3 &=&  &&  c_{31} \displaystyle\frac{v^4}{r^{13}} & +
& c_{32} \displaystyle\frac{v^6}{r^{17}}& + &
c_{33} \displaystyle\frac{v^8}{r^{21}} & +\ \ldots
\end{array}
\label{mmseries}
\end{equation}
In writing this expansion we have used the fact that the supersymmetry
algebra with~16 supercharges prevents renormalization of the
coefficient of $v^2$, and that the expansion must be even in $v$ by
time-reversal invariance.

Now let us consider the supergravity prediction.  We will study  the
scattering of gravitons of momenta $p_- = N_1 / R$, $N_2 / R$,
with
$N_1$ large enough that the first graviton can be considered as a classical
source for the gravitational field.  Ultimately,  to understand the full form
of the supergravity amplitude we will need to develop the Feynman rules for
supergravity with lightlike compactification, but we leave that for future
work.

Our conventions are $x^{\pm} = x^{11} \pm t$, and the time parameter of
the light-cone quantization is $\tau = \frac{1}{2} x^+$.  These
are chosen so that the large-$N$ limit of the lightlike quantization is
consistent with the large-$N$ limit of spacelike compactification.
Note that at large $N$ all particles are moving approximately along
lines of $\delta x^{11} = \delta t$, so that with $x^{-} = x^{11} - t$
the periodicity is $2\pi R$ both in $x^-$ and $x^{11}$.  Also, $\delta
\tau = \frac{1}{2}(\delta x^{11} + \delta t) \sim \delta t$ along any
world-line.  Finally, $p_-$ is positive.

The source graviton is taken to have vanishing transverse velocity.
Its world-line is $x^- = x^i = 0$, and it produces the Aichelburg-Sexl
metric \cite{aichse}
\begin{equation}
G_{\mu\nu} = \eta_{\mu\nu} + h_{\mu\nu}\ ,
\end{equation}
where the only nonvanishing component of $h_{\mu\nu}$ is
\begin{equation}
h_{--}\ =\ \frac{2 \kappa_{11}^2 p_-}{7 \omega_8 r^7} \delta(x^-)
\ =\ \frac{15 \pi N_1}{ R M^9 r^7} \delta(x^-). \label{hmm}
\end{equation}
Here $\kappa_{11}^2 = 16 \pi^5/ M^{9}$ (see ref.~\cite{bercor}, for example)
and
$\omega_8$ is the volume of $S_8$.
This metric can be thought of as obtained from the Schwarzchild metric by
taking the limit of infinite boost in the $+$ direction while the mass is
taken to zero; the latter accounts for the absence of higher-order terms
in
$1/r$ or $N_1$. A more detailed derivation of this metric can be found
in the appendix.
The source graviton is in a state of definite $p_-$ and so we
average over the $x^- \in (0,2\pi R)$ direction to give
\begin{equation}
h_{--}\ =\ \frac{15 N_1}{2 R^2 M^9 r^7} \ . \label{hmmav}
\end{equation}

For the action of the `probe' graviton in this field we use the following
trick.  Begin with the action for a massive scalar (spin effects
 fall more rapidly
with $r$) in eleven dimensions
\begin{eqnarray}
S &=& - m \int d\tau \,(-G_{\mu\nu}\dot x^\mu \dot x^\nu)^{1/2}
\nonumber\\
&=& - m \int d\tau \, \left( -2\dot x^- - v^2 - h_{--} \dot x^- \dot x^-
\right)^{1/2}\ ,
\end{eqnarray}
where we have used the form of the Aichelburg-Sexl metric.  A
dot denotes $\partial_\tau $
and $v^2 = \dot x^i \dot x^i$.
This action vanishes if we take $m \to 0$ with
fixed  velocities, but for the process being considered here it is $p_-$
that is to be fixed.  We therefore carry out a Legendre transformation on
$x^-$:
\begin{equation}
p_- = m\, \frac{1 + h_{--} \dot x^-}{\left( -2\dot x^- - v^2 - h_{--}
\dot x^- \dot x^- \right)^{1/2}}\ . \label{legend}
\end{equation}
The appropriate Lagrangian for $x^i$ at fixed $p_-$ is (minus) the
Routhian,
\begin{equation}
{\cal L}'(p_-)\ =\ - {\cal R}(p_-)\ =\ {\cal L} - p_- \dot x^-(p_-) \
.\label{routh}
\end{equation}
Eq.~(\ref{legend}) determines $\dot x^- (p_-)$; it is
convenient before solving to take the limit $m \to 0$, where it reduces
to $G_{\mu\nu}\dot x^\mu \dot x^\nu=0$.  Then
\begin{equation}
\dot x^- = \frac{\sqrt{1 - h_{--} v^2} - 1}{h_{--}}\ .
\end{equation}
In the $m \to 0$ limit at fixed $p_-$ the effective Lagrangian becomes
\begin{eqnarray}
{\cal L}' &\to & -p_- \dot x^-
\nonumber\\
&=& p_- \left\{ \frac{v^2}{2} + \frac{h_{--} v^4}{8} +
\frac{h_{--}^2 v^6 }{16} + O(h_{--}^3 v^8) \right\}
\nonumber\\
&=& \frac{N_2}{2R} v^2 + \frac{15}{16}\frac{N_1 N_2}{R^3 M^9
}\frac{v^4}{r^7} + \frac{225}{64 }
\frac{ N_1^2 N_2}{ R^5 M^{18} }\frac{v^6}{r^{14}}  + O\left(
\frac{v^8}{r^{21}}\right)\ .
\label{routhtwo}
\end{eqnarray}

The $r$ and $v$ dependences match the diagonal terms in the
series~(\ref{mmseries}), and the $N$-dependences are consistent with the
leading large-$N$ behavior $N^{L+1}$.
The $v^4/r^7$ agrees with the one-loop matrix model amplitude, as asserted in
ref.~\cite{bfss} and worked out in detail in ref.~\cite{bercor}.
The two-loop calculation in ref.~\cite{beckers} extended to $v^6$ gives for
$SU(2)$ the value
\begin{equation}
\frac{225}{32} \frac{1}{R^5 M^{18}}\frac{v^6}{r^{14}}\ . \label{essyootoo}
\end{equation}
In~\cite{beckers},
$R M^3$ was implicitly set to one, but we have restored it by dimensional
analysis and used the relation $g = 2R$ for $g$ defined in
ref.~\cite{beckers}, as follows from the tree level term
in~(\ref{routhtwo}).
The separate contributions of the various two-loop graphs are
given in the table.
\begin{table}
\begin{center}
\begin{tabular}{clclclcl}
$(a)_1 $& $\displaystyle +\frac{287481}{28672}$ &
$(b)_3 $& $\displaystyle -\frac{4717523}{229376}$ &
$(c)_1 $& $\displaystyle +\frac{107251}{114688}$ &
$(d)_2 $& $0$ \\[9pt]
$(a)_2 $& $\displaystyle -\frac{27519}{7168}$ &
$(b)_4 $& $\displaystyle +\frac{16965}{4096}$ &
$(c)_2 $& $\displaystyle +\frac{892261}{688128}$ &
$(d)_3 $& $\displaystyle +\frac{4615}{672}$ \\[9pt]
$(b)_1 $& $\displaystyle -\frac{2366913}{114688}$ &
$(b)_5 $& $\displaystyle +\frac{13311}{4096}$ &
$(c)_3 $& $\displaystyle +\frac{231}{1024}$ &
$(d)_4 $& $\displaystyle +\frac{7995}{224}$\\[9pt]
\cline{7-8}
$(b)_2 $& $\displaystyle +\frac{31595}{14336}$ &
$(b)_6 $& $\displaystyle +\frac{315}{2048}$ &
$(d)_1 $& $\displaystyle -\frac{698165}{43008}$  &
sum&$\displaystyle +\frac{225}{64}^{\vphantom h}$
\end{tabular}
\end{center}
\caption{Coefficients of $ gv^6/r^{14}$.  Graphs are labeled as in
ref.~[5]. We have included factors $1/2$ for diagrams involving two
cubic vertices of the same type directly in this table.}
\end{table}
The $N$-dependence can be reconstructed as follows.  In double-line notation
every graph involves three index loops, and so is of order~$N^3$.  Terms
proportional to $N_1^3$ or $N_2^3$ would only involve one block (graviton)
and so could not depend on $r$.  Symmetry under interchange of $1$ and $2$
thus determines that the $SU(2)$ result~(\ref{essyootoo}) is multiplied by
\begin{equation}
\frac{N_1 N_2^2 + N_1^2 N_2}{2}\ ,
\end{equation}
in agreement with the supergravity result~(\ref{routhtwo}) for the term of
interest.  

Note that we have not distinguished radial and transverse
velocities.  Any term proportional to the radial velocity is equivalent by
parts to a term involving the acceleration.  All matrix theory calculations to
date have considered straight-line motion and so are insensitive to such
terms.  Thus, we write $v^2$ with the understanding that only the transverse
part is relevant.

It is difficult to be certain which of the many tests of matrix theory
actually test that conjecture and not just the weaker and less
controversial assumption that the IIA string has an eleven-dimensional limit.
In the present case the numerical agreement is impressive.   Moreover, it is
difficult to see how supersymmetry alone would determine the
normalization of the $v^6/r^{14}$ term in the supersymmetric quantum
mechanics effective action, suggesting that an additional structure
(eleven-dimensional Lorentz invariance) is present. A die-hard skeptic
might still argue as follows.  A `normal' supersymmetric invariant,
obtained from a multiple commutator with all sixteen supercharges (the
analog of an integral over all of superspace), would be at least of order
$v^8$.  The $v^6$ term is therefore `chiral' and so might be constrained by
nonrenormalization theorems.  Then one could continue from the eleven
dimensional supergravity limit where one calculation is valid, to the IIA
string limit where the other calculation is valid, and the answers must agree
independent of the matrix theory conjecture.  But the skeptic is not willing
to bet that the $v^8 /r^{21}$ term, a three-loop effect, will show a
discrepancy.

It is interesting to consider higher corrections in the supergravity theory.
The $v^6/r^{14}$ term can be though of as arising from the graph of figure~1a,
with a second order coupling to the probe.\footnote{Note that this is second
order in a first quantized description of the probe.  This does not correspond
directly to second order in a field theory action.}
\begin{figure}
\begin{center}
\leavevmode
\epsfbox{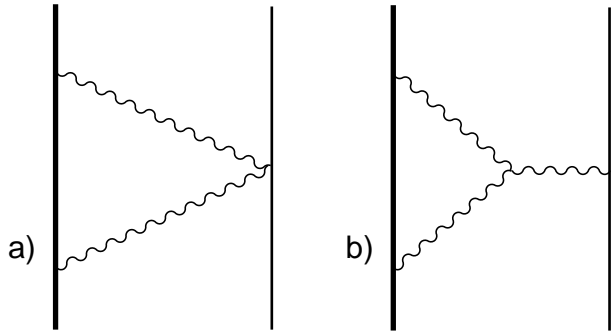}
\end{center}
\caption[]{a) Graphical representation of probe graviton (thin straight line)
interacting with the metric of source graviton (heavy straight line) at second
order.  b) Vanishing nonlinear correction to the metric of the source.}
\end{figure}
Figure~1b would
represent a nonlinear correction to the metric~(\ref{hmmav}), which as we have
noted is absent.  The ladder graph is second order in the effective Lagrangian
and the crossed ladder is absent in the light-cone frame. It appears that each
graviton coupling to the source brings at least an
$r^7$ from the field~(\ref{hmmav}), so that the leading large-$r$ behavior at
order
$N_1^k$ would be the diagonal $k$-loop term that we have considered.

Dimensionally, higher derivative operators in the low energy supergravity
theory bring in additional powers of $v$ and $1/r$ and so correspond to
matrix model amplitudes that lie, for a given power of $N_1$, to the right of
and/or below $N_1^k v^{2k+2} / r^{7k}$ in the series~(\ref{mmseries}). 
However, higher-order {\it local} curvature invariants
$R^4 + R^6 + ...$ in the $D=11$ supergravity action  are not expected to
change the low-energy scattering of
two gravitons  when one of them has large
$p_-$, i.e. when it can be treated as a source for the gravitational field.
The reason is that (in contrast to the Schwarzschild solution, for example)
the corresponding  plane-fronted wave background (\ref{hmmav}) is
not modified  by $R^n$
 corrections to the action: according to the standard argument (see, e.g.,
\cite{gary}),  the existence of a covariantly constant
null Killing vector implies
the vanishing of all second rank tensors constructed
out of curvature and the metric  except the  Ricci one
(equivalently, corrections  to Schwarzschild  disappear
in the  infinite boost, zero mass limit).
Supergravity loop effects, being weak at low energy,
should also lie to the right of and below the $N_1^k v^{2k+2} / r^{7k}$ term.
Terms below the diagonal that arise in this way are subleading in
$N_1$ for the given number of loops.\footnote{This has also been noted by W.
Fischler and L. Susskind~\cite{susspriv}.}  It would be interesting to relate
these supergravity effects to the matrix model, even at one matrix model loop
where the whole series is known~\cite{oneloop}. In passing we would like to
mention the observation that the coefficient
$c_{12}$ of the next higher one-loop term $v^6/r^{11}$ actually vanishes.

On the matrix model side there is the
important complication of bound state effects.\footnote{We would like to thank
David Gross for raising this issue.  See also ref.~\cite{ggr}.} Matrix theory
scattering calculations to date have treated the zero-branes in a bound state
as being coincident with zero relative velocity.  Note, however, that a term
which is dimensionally of order
$v^8$ can have the structure $v_1^2 v_2^6$ and even with the center-of-mass
$v_1$ vanishing can generate a
$v_2^6$ term proportional to the expectation value of the relative $v^2$
in the bound state.  This would not affect the present calculation because
all $v^8$ terms fall off more rapidly in $r$, but to determine some higher
terms one needs an understanding of the bound state.  One must also consider
recoil, interactions causing the gravitons to deviate from a straight line.  To
the order we are working we believe that this corresponds to omitting the
one-particle-reducible two-loop graphs, but at higher order it may be necessary
to separate the light and heavy matrix model degrees of freedom in a more
systematic way.

It is interesting to repeat the derivation of the Routhian for scattering at
fixed spacelike momentum $p_{11}$.  Here  we have  ($\tau=t$)
\begin{equation}
S = - m \int dt\, \left\{ 1 - (\dot x^{11})^2 -
 v^2 -  h_{--} (\dot x^{11} -
1)^2 \right\}^{1/2} \ .
\end{equation}
Then one finds
\begin{eqnarray}
{\cal L}' &=& -p_{11} \dot x^{11}
\nonumber\\
&=& -p_{11}  \bigg[ 1 +
\frac{\sqrt{1 - ( 1 +  h_{--}) v^2} - 1}{1 +  h_{--}}\bigg] \ .
\label{routh3}
\end{eqnarray}
Where the earlier Routhian (\ref{routhtwo})
 had only terms of order $v^2 (v^2/r^7)^k$, this now
has higher velocity corrections, a double series $v^{2 + 2l} (v^2/r^7)^k$.
Spacelike compactification of M theory gives the IIA string
 theory, and eq.(\ref{routh3}) is precisely the
action for interaction of two D0-branes via classical supergravity.  This is
the more complicated expansion considered in ref.~\cite{ballar}, but we see
that it has no direct relevance to finite $N$ matrix theory.  We emphasize
that the result~(\ref{routhtwo}) is fully relativistic.

It is curious that the null and timelike Lagrangians
(\ref{routhtwo})  and (\ref{routh3})
 are related by the simple
substitution $ h_{--} \to 1 +  h_{--}$
(the transverse velocities  are in direct correspondence because of our
conventions, as noted earlier).  To better understand the formal relation
between the two cases, note that just as (\ref{routh3}) is
essentially  the
  Lagrangian  for  a  D0-brane probe
moving in a D0-brane source background,
(\ref{routhtwo})  can be interpreted as a D0-brane probe
Lagrangian in a $D=10$  background
resulting from   reducing the  $D=11$ plane wave \
$
ds^2_{11} = dx^+ dx^- + h_{--} dx^- dx^-  + dx^i dx^i  ,
 \ \  h_{--} = { Q \over r^7} , \
$
along  the null $x^-$ direction\footnote{More precisely, the direction  $x^-$
is  null in flat space but is space-like in the curved plane wave
background.}
   instead of  the spatial
$x^{11}$ direction. While  the reduction
along $x^{11}$  gives the standard 0-brane background,
the reduction along $x^-$ produces  the following
$D=10$  (string-frame) metric, dilaton and 1-form field
\begin{equation}
 ds^2_{10 } =  - h_{--}^{-1/2} d\tau^2  +  h_{--}^{1/2} dx^i dx^i \ , \ \ \
e^{\phi} = h_{--}^{3/4},  \ \ \ A= - h_{--}^{-1} d\tau \ ,
\label{null}
\end{equation}
where  $\tau= \frac{1}{2}  x^+$.
This becomes the usual
0-brane solution   if  $\tau \to t $ and $h_{--} \to H=1+h_{--}$
 (and  $A\to A + dt$).
This relation is implied by
the structure of the  $D=11$ plane wave metric (in particular,
it remains invariant under
 $x^+ \to x^+ - x^- = 2t$  and $h_{--} \to h_{--} + 1$).

Thus  (\ref{null}) is formally the same as
the {\it short-distance}  (or `near-horizon') limit of the 0-brane
background: then $h_{--} \gg 1$ so that $H=1 +h_{--} \approx h_{--}$.
Equivalently, it may be viewed as a
   large
charge $Q\sim N g $
 or {\it large $N$}  (but  {\it fixed} distance  $r$)
limit of the 0-brane solution.
The  fact that the  two actions are
formally related  by $h_{--} \to 1 + h_{--}$ implies that
large $r$, small $v$  expansion of the  first   action
is  simply the leading part of  the expansion of the second action.

As was already mentioned above,  it is the
 `null reduction'  action
that  is  in direct correspondence with the
 matrix theory results for finite $N$.
Remarkably, this conclusion extends also to more complicated cases
of graviton scattering off M-branes discussed in \cite{morescat}.
Again, the  supergravity potentials corresponding to
the `fixed $p_-$' case can be obtained from the relevant
D-brane probe  actions in the  $D=10$  backgrounds
 following upon  reduction along the `null' direction $x^-$.
These actions  are found  from  the `fixed $p_{11}$' actions by
 replacing   the 0-brane harmonic function $H$  by its `short-distance' (or
large $N$) limit  $H-1=h_{--}$.  The resulting
  long-distance  interaction
potentials (containing in general
both static and velocity-dependent  terms like
$V= {1\over r^n} (a + b v^2 + cv^4)+ O({1\over r^{2n}})$)
 are then in precise agreement with the
 one-loop matrix model potentials with no need
to  assume that the number of 0-branes $N$ is  large
as was done in \cite{morescat}
 to  be able  to ignore additional  subleading terms present
in the fixed $p_{11}$  picture.
 This provides another test of  the  proposal of ref.~\cite{dlqm}.

Finally, let us note that from another point of view, discrete light-cone
quantization can be regarded as a limit of spacelike compactification as
follows \cite{helpol}.
The null direction has zero invariant length, so by a boost should be
related to the $R_{11} \to 0$ limit.  The naive $R_{11}\to 0$
limit is simply dimensional reduction to the
$p_{11}R_{11} = 0$ sector.  Here one takes instead the $p_{11}R_{11} = N$
sector, subtracts the overall $N/R_{11}$ and rescales to
\begin{equation}
H_{\it eff} =
\frac{H - N/R_{11}}{R_{11}}\ .
\end{equation}
at fixed momentum $p^i$.
Noting that $v^i = O(R_{11} p^i)$ and $h_{--}
= O(R_{11}^{-2})$, this yields eq.~(\ref{routhtwo}) from eq.~(\ref{routh3}).


\subsection*{Acknowledgments}
We would like to thank David Gross, Simeon Hellerman, and Lenny Susskind for
helpful conversations.  This work was supported in part by NSF grants
PHY91-16964 and PHY94-07194,  DOE grant DOE-91ER4061,
 PPARC and  the EC TMR grant ERBFMRX-CT96-0045.

\subsection*{Appendix}
In this appendix we would like to derive the form of the Aichelburg-Sexl
metric (\ref{hmm}). This
follows closely the original derivation of \cite{aichse}. Start with the
Einstein field equations
\begin{equation}
R_{\mu \nu}-\frac{1}{2} G_{\mu \nu } R= \kappa^2_{11} T_{\mu \nu },
\end{equation}
and approximate
\begin{equation}
G_{\mu \nu }=\eta_{\mu \nu } +h_{\mu \nu },
\end{equation}
where $(h_{\mu \nu})^2 \approx 0$. This gives the linearized
field equations:
\begin{equation}
(\partial_t^2 -\triangle) \psi^{\mu \nu }=
2\kappa^2_{11} T^{\mu \nu } \label{lifieq},
\end{equation}
where:
\begin{equation}
\psi^{\mu \nu }=h^{\mu \nu } -\frac{1}{2} \eta^{\mu \nu } {h_{\lambda}}^
{\lambda}.
\end{equation}
For a massless particle moving in $x_{11}$ direction
with the velocity of light the energy momentum tensor is
\begin{equation}
T^{\mu \nu }=p_- \delta(x^-) \delta(x_{\perp}) s^{\mu} s^{\nu}, \label{tik}
\end{equation}
where $s^{\mu}=\delta_0^{\mu}+\delta_{11}^{\mu} $ and $\delta (x_{\perp})=
\prod_{i=1}^{9} \delta (x_i)$. Inserting (\ref{tik})
in (\ref{lifieq}) gives a determining equation for $\psi^{\mu \nu }$. To
solve it make the ansatz:
\begin{equation}
\psi^{\mu \nu}= 2 \kappa_{11}^2 p_-\delta(x^-) G_9(x_{\perp}) s^{\mu} s^{\nu}.
\end{equation}
Then $G_9$ satisfies 9-dimensional Poisson equation:
\begin{equation}
\triangle G_9(x_{\perp})+\delta (x_{\perp})=0.
\end{equation}
The solution is given by
\begin{equation}
G_9(x_{\perp})= \frac{15}{2(2 \pi)^4} \frac{1}{r^7}.
\end{equation}
Therefore
\begin{equation}
\psi^{\mu \nu }=\frac{ 15\pi N_1 }{ R M^9 r^7 }\delta(x^-) s^{\mu} s^{\nu}  .
\end{equation}
This determines the form of the fluctuation of the metric.
The only nonvanishing component of $h_{\mu \nu}$ is $h_{--}$ with the result
(\ref{hmm}).

\end{document}